\documentclass[aps,prl,twocolumn,amsmath,amssymb,nofootinbib,superscriptaddress,floatfix]{revtex4-1}

\usepackage{amsmath}
\usepackage{amssymb}
\usepackage{amsthm}

\usepackage[dvips]{color} 

\usepackage{graphicx}% Include figure files
\usepackage{dcolumn} % Align table columns on decimal point
\usepackage{bm} % bold math

\usepackage[hypertex]{hyperref}
\usepackage{longtable}
\usepackage{multirow} %Tables with joned rows and columns
\usepackage{ulem}   % to strike things out
\newcommand{\bs}[1]{{\boldsymbol{#1}}}

\begin{document}

\title{Chain of Majorana States from Superconducting Dirac Fermions at a Magnetic Domain Wall
}

\author{Titus Neupert} 
\affiliation{
Condensed Matter Theory Group, 
Paul Scherrer Institute, CH-5232 Villigen PSI,
Switzerland
            } 
\affiliation{
Condensed Matter Theory Laboratory,
RIKEN, Wako, Saitama 351-0198, Japan
            }
\author{Shigeki Onoda} 
\affiliation{
Condensed Matter Theory Laboratory,
RIKEN, Wako, Saitama 351-0198, Japan
            } 
\author{Akira Furusaki} 
\affiliation{
Condensed Matter Theory Laboratory,
RIKEN, Wako, Saitama 351-0198, Japan
            } 

\date{\today}

\begin{abstract}
We study theoretically a strongly type-II $s$-wave superconducting state
of two-dimensional Dirac fermions in proximity to
a ferromagnet having in-plane magnetization.
It is shown that a magnetic domain wall can host a chain of
equally spaced vortices in the superconducting order parameter,
each of which binds a Majorana-fermion state.
The overlap integral of neighboring Majorana states is sensitive to
the position of the chemical potential of the Dirac fermions.
Thermal transport and scanning tunneling microscopy
experiments to probe 
the Majorana fermions are discussed.
\end{abstract}

\maketitle

Majorana fermions are hypothetical particles that are their own
antiparticles, as originally proposed to describe neutrinos~\cite{Majorana37}.
Because of the non-Abelian statistics as Ising anyons in two spatial
dimensions~\cite{Nayak08} and their potential application to a fault-tolerant
quantum computation~\cite{Nayak08,Kitaev03,Tewari07},
the Majorana fermions have attracted revived interest in condensed-matter
physics~\cite{Wilczek09}.
Theoretically, they appear as zero-energy states bound to vortices in
topological superconductors~\cite{Read00},
including  $\nu=5/2$ fractional quantum-Hall states~\cite{Nayak08},
$s$-wave superconductors of two-dimensional (2D)
Dirac fermions~\cite{Jackiw81}, and
chiral $p$-wave superconductors~\cite{Read00,Ivanov01}.
However, it remains a challenge to realize and manipulate
Majorana fermions in a practically detectable and controllable manner. 

Recently, there have been several proposals for realizing
Majorana fermions as the edge modes at heterostructure interfaces.
It has been argued that the superconductivity induced by the proximity
effect on the surface of a three-dimensional (3D) topological
insulator (TI)~\cite{Fu08, Fu09, Akhmerov09} can host
Majorana modes propagating along the edge of the surface.
When the superconducting region is surrounded by two magnets
having opposite out-of-plane magnetizations,
two charge-neutral Majorana edge modes are combined into
a charged Dirac fermion, which can be probed with charge transport.
A Rashba-semiconductor thin film sandwiched by an insulating magnet
having an out-of-plane magnetization and an $s$-wave superconductor
has also been proposed to produce a chiral Majorana edge
mode~\cite{Sato09,Sau10-1,Sau10-2,Alicea10}. 

\begin{figure}[b]
\includegraphics[angle=0,scale=0.3]{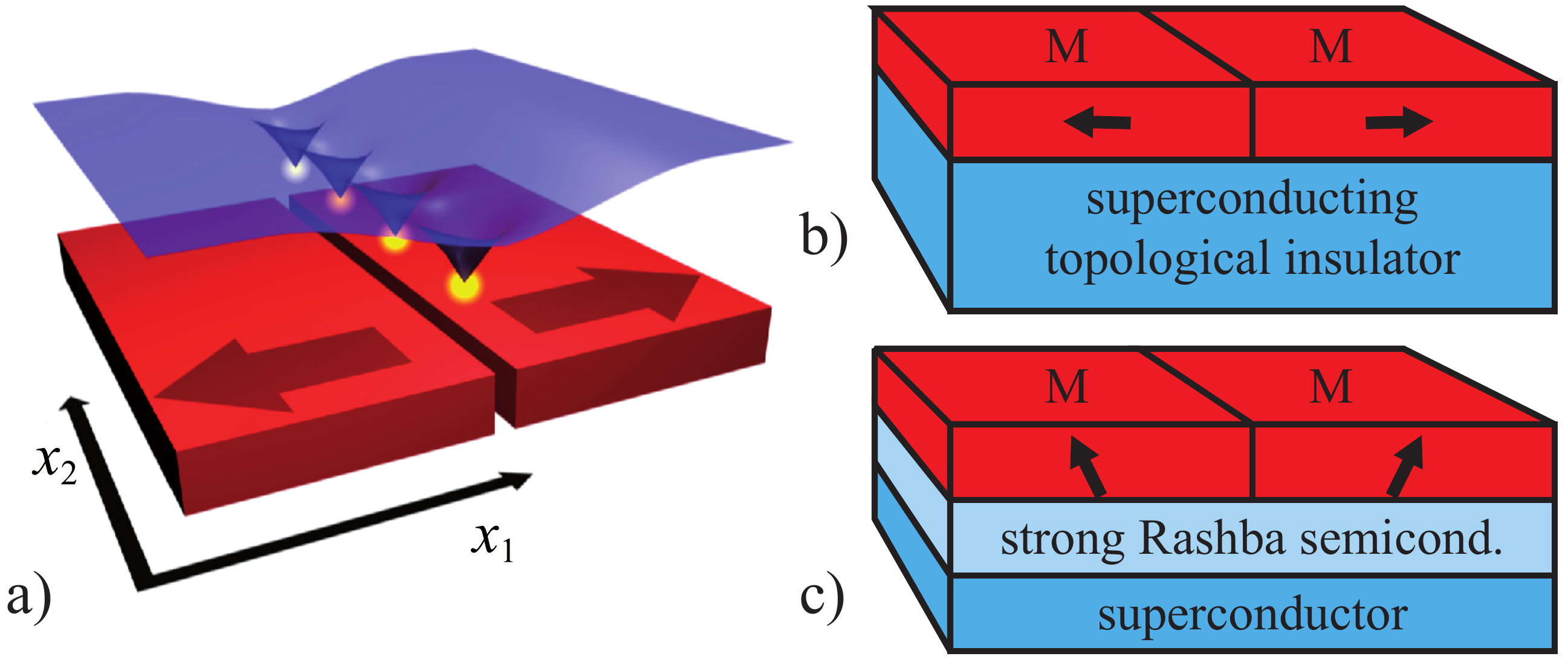}
\caption{(Color online)
(a)
Schematic picture of the modulus of the superconducting order
parameter (thin surface)
in either of the two proposed heterostructures shown in (b) and (c).
Along the boundary between two domains of a ferromagnetic insulator (M)
with opposite in-plane magnetization
(bold arrows), a chain of vortices emerges.
A single Majorana fermion is bound to each vortex (dots).
         }
\label{fig:Heterostructure}
\end{figure}

In this Letter we show theoretically that
a magnetic domain wall in an insulating ferromagnet attached to
a 2D strongly type-II $s$-wave Dirac-fermion superconductor can produce
a chain of vortices [Fig.~\ref{fig:Heterostructure}(a)],
each accommodating a Majorana zero mode~\cite{Kitaev03},
when the magnetization points to the out-of-plane direction
at the head-to-head magnetic domain wall.
A hybridization of Majorana fermions at neighboring vortices
can be tuned by the chemical potential of the Dirac-fermion superconductor,
which can be probed with thermal transport
and scanning tunneling microscopy (STM) experiments.
We also propose two heterostructures that can realize such electronic states:
(A) an interface of an insulating ferromagnetic thin film and
a bulk $s$-wave superconductor whose normal state
is a 3D TI [Fig.~\ref{fig:Heterostructure}(b)]
and (B) a device structure with stacked $s$-wave superconductor,
Rashba semiconductor, and insulating magnetic film
[Fig.~\ref{fig:Heterostructure}(c)].

We consider a 2D continuum BCS-Dirac Hamiltonian with
a single Dirac cone in the single-particle dispersion,
\begin{eqnarray}
H_0\!\!&=&\!\!
\int\! d^2\bs{x} \, \psi^\dagger(\bs{x})
\{
\left[iv\bs{D}\times\bs{e}_3-\bs{B}(\bs{x})\right]\cdot\bs{\sigma}
-\mu\sigma_0
\}
\psi(\bs{x})
\nonumber\\
&&\!\!
+\int\! d^2\bs{x} \left\{\Delta(\bs{x})\psi^\dagger(\bs{x})i\sigma_2
\!\left[\psi^\dagger(\bs{x})\right]^T+\text{H.c.}\right\},
\label{eq: dispersion}
\end{eqnarray}

under the Zeeman field $\bs{B}(\bs{x})$,
i.e.,
an exchange field
coming from a ferromagnet.
Here, $v$ is the Dirac-fermion velocity,
$\mu$ the chemical potential, and
the covariant derivative
$\bs{D}=\nabla-ie\bs{A}$
with $\bs{A}$ being the vector potential.
The spinors
$\psi^\dagger(\bs{x})=(c^\dagger_{\bs{x}\uparrow},c^\dagger_{\bs{x}\downarrow})$
and
$\psi(\bm{x})=(c_{\bm{x}\uparrow}^{},c_{\bm{x}\downarrow}^{})$
create and annihilate fermions at position $\bs{x}$ with charge $e$,
respectively.
We use $\bs{\sigma}=(\sigma_1,\sigma_2,\sigma_3)$ for the Pauli matrices
as well as $\sigma_0$ for the 2$\times$2 unit matrix.
The unit vector $\bs{e}_3$ is normal to the 2D plane,
while 
$\bs{x}\equiv x_1\bs{e}_1+x_2\bs{e}_2$ lies in the plane.
In the absence of the magnetic field $\bs{B}(\bs{x})=0$, the superconducting order parameter $\Delta(\bs{x})$ is assumed to take
a uniform constant $\Delta_0$, which we choose real and positive. 
Henceforth, we assume strong type-II superconductivity
so that the orbital coupling to the vector potential $\bm{A}$
can be ignored
in the analysis of $\Delta(\bs{x})$.

\begin{figure}[b]
\includegraphics[angle=0,scale=0.3]{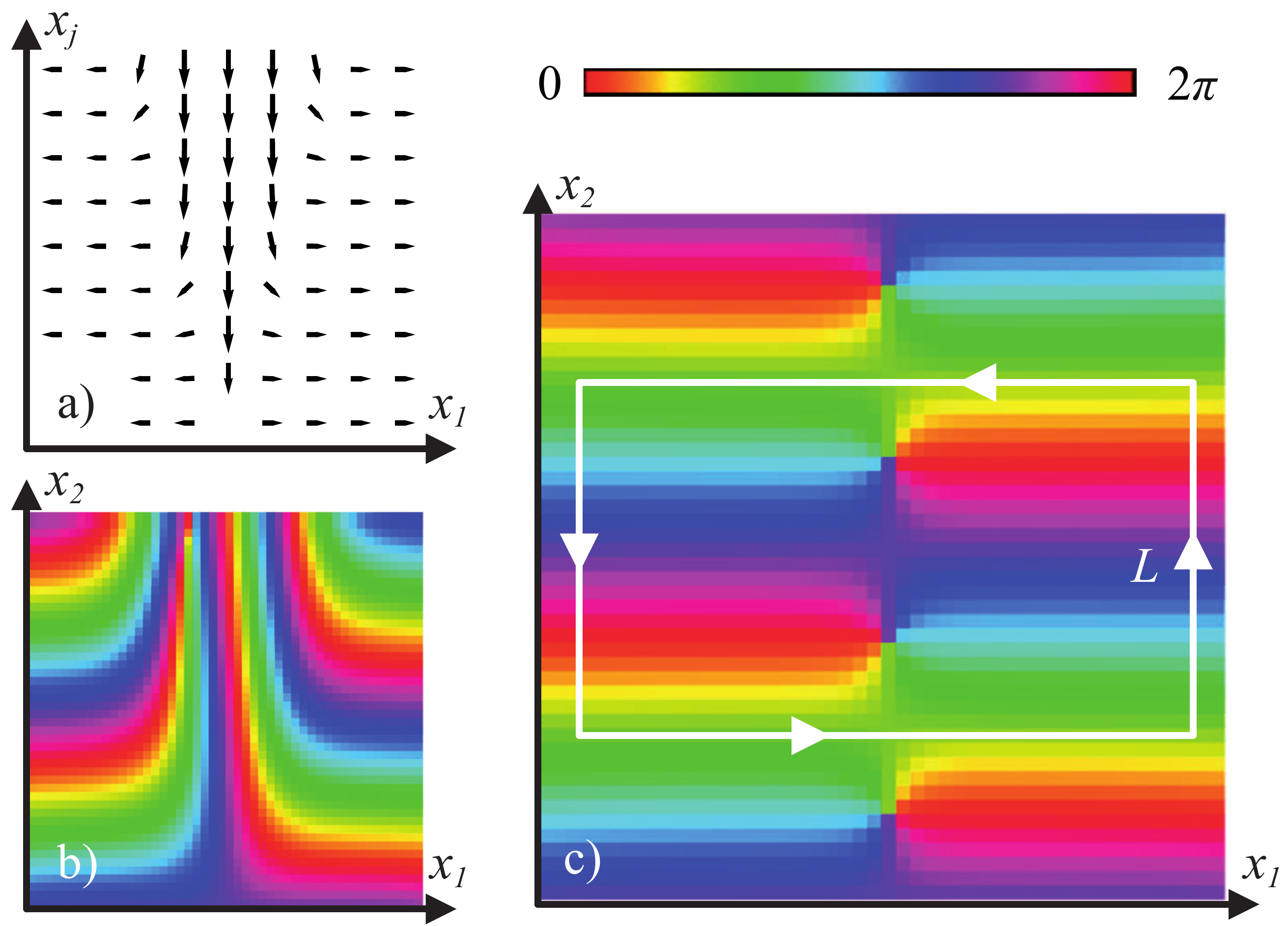}
\caption{(Color online)
(a) Magnetic moment of the ferromagnet around the domain walls of types (i) and (ii). 
This structure continues in the other spatial direction. 
(b) and (c)
Color maps of the phase of the superconducting order parameter $\Delta(\bs{x})$
in our numerical solutions of the GL model Eq.~\eqref{eq:Free energy}
on a grid of 50$\times$50 sites for the types (i) and (ii)
of head-to-head domain walls, respectively.
They have been obtained at the criticality by ignoring the quartic term.
In the $\bs{e}_2$ direction, open boundary conditions and periodic
boundary conditions were chosen for cases (i) and (ii), respectively.
Fixed boundary conditions resembling the solutions deep inside
each domain were imposed in the $\bs{e}_1$ direction.
The domain-wall parameters were set to $l=6$ lattice sites and $s=1/4$
while $B/v$ was chosen such that two periods of
the phase modulation fit into the grid.
A variation in the model parameters $l$, $s$ and $|B/v|$ does not
change the result qualitatively.
        }
\label{fig:Setup}
\end{figure}

Here, we introduce a head-to-head domain wall that extends along
the $\bs{e}_2$-direction [Fig.~\ref{fig:Setup}(a)]. 
Away from the domain wall, the Zeeman field is given by
$\bs{B}\to\pm B\bs{e}_1$
for $x_1\to\pm\infty$, 
respectively.
Therefore, it can be approximated with a uniform
in-plane Zeeman field $\pm B\bs{e}_1$.
This shifts the Dirac cone and therefore
causes a finite center-of-mass momentum
$\bs{q}=\pm q\bs{e}_2$ of the spin-singlet Cooper pairs with
$q=2B/v$~\cite{Santos10}.
As a consequence, the superconducting order parameter $\Delta$
will acquire a spatial modulation 
$\Delta(\bs{x})=\Delta^{\ }_0e^{ i\bs{q}\cdot\bs{x}}$
for $x_1\to\pm\infty$.

Alternatively, the spatial modulation of $\Delta(\bs{x})$
by the in-plane Zeeman field can be described
using the Ginzburg-Landau (GL) free energy,
\begin{equation}
\begin{split}
F=&\int d^2\bs{x}\left[
\alpha |\Delta(\bs{x})|^2+\gamma|\mathcal{D}_\Delta\Delta(\bs{x})|^2
 +u|\Delta(\bs{x})|^4
\right],
\end{split}
\label{eq:Free energy}
\end{equation}
where $\mathcal{D}_\Delta=\nabla-2i(e\bs{A}+v^{-1}\mathcal{A})$
and we introduced
the real parameters $\alpha$, $\gamma>0$, and $u>0$.
The linear coupling to the Zeeman field, written in terms of the fictitious 
vector potential $\mathcal{A}:=\bs{e}_3\times\bs{B}$,
is allowed due to the broken mirror
symmetry about the 2D plane.
Away from the domain wall,
the mean-field solution in the strongly type-II limit can indeed
be obtained as 
$\Delta(\bs{x})=\Delta^{\ }_0e^{ i\bs{q}\cdot\bs{x}}$
with $\Delta_0^{\ }=\sqrt{|\alpha|/2u}$ for $\alpha<0$.
We employ this GL theory
to determine a spatial profile of $\Delta(\bs{x})$
in the domain-wall region where the Zeeman field $\bs{B}$ varies with $\bs{x}$.

In fact, the profile of $\Delta(\bs{x})$ strongly depends on
whether the magnetic moment of the attached ferromagnetic film changes its direction along the domain wall
(i) within the plane
as in stripes of magnetic thin films~\cite{Klaeui08} 
or 
(ii) towards the out-of-plane direction
as in magnetic thin films with large magnetic anisotropy~\cite{Kittel46}.
The magnetic moment
near these domain walls can be modeled
as a divergence-free function,
\begin{equation}
\bs{M}(\bs{x},x_3)=B\sum_{\lambda=\pm s}
\frac{\bs{e}_1-\lambda\bs{e}_j}{2}
\tanh\left(\frac{\lambda x_1+x_j}{\lambda l}\right),
\label{eq:B}
\end{equation}
with $j=2$ for the case (i) and $j=3$ for the case (ii),
where $l$ and $s$ control the characteristic width of the domain wall.
We assume that the Zeeman field $\bs{B}$ acting on the Dirac-fermion superconductor at the interface $x_3=0$ is given by $\bs{B}(\bs{x})=\bs{M}(\bs{x},0)$.
A key observation here is that the out-of-plane component $B_3(\bs{x})$ 
vanishes. 
Then, this in-plane Zeeman field induces a fictitious magnetic field defined by
$\mathcal{B}:=\nabla\times\mathcal{A}$, which becomes $\mathcal{B}=0$ and 
$\mathcal{B}=\bs{e}_3(B/l)\,\text{sech}^2{(x_1/l)}$
for cases (i) and (ii), respectively.
In analogy to the vortex-lattice phase of a type-II superconductor,
we would thus expect 
to obtain vortices at the domain wall only for case (ii).
This is confirmed by the numerical solutions
of the phase of the superconducting order parameter $\Delta(\bs{x})$
for the cases (i) and (ii), presented in Figs.~\ref{fig:Setup}(b) and (c),
respectively.
For case (i), the contour of the phase of $\Delta(\bs{x})$
follows the magnetic flux $\bs{B}$ and hence no vortices emerge.
For case (ii), a chain of vortices in $\Delta(\bs{x})$ appears
along the domain wall.
All the vortices have the same winding number $+1$ or $-1$,
where the sign is determined by the sign of $B/v$.

The emergence of the vortices in case (ii) can also be understood from
the integration of the phase of 
$\Delta(\bs{x})$
deep inside the two domains, i.e.,
along the contour drawn in Fig.~\ref{fig:Setup}(c).
Provided 
$\Delta(\bs{x})$
does not vanish along the contour,
the flux enclosed by this contour takes an integer multiple of the
superconducting flux quantum $\Phi_0$, that monotonically changes by 1
each time the contour length along the domain wall is increased by $\pi/q$.
It indicates that the intervortex distance is given by $\pi/q$.
In Fig.~\ref{fig:Setup}(c) we observe that 
$\Delta(\bs{x})$
changes sign
when translating it by the separation of neighboring vortices
$\bs{x}_q=(0,\pi/q)$, i.e.,
$\Delta(\bs{x})=-\Delta(\bs{x}+\bs{x}_q),\ \forall\bs{x}$.

Having determined the spatial structure of the order parameter,
we will now discuss how Majorana fermions emerge.
An \emph{isolated} vortex in the superconducting order parameter
comprises a Majorana fermion as a bound state at precisely zero energy.
However, in the vortex chain, these bound states can potentially
couple to each other and form an energy band around zero energy.
To study this, we will construct an effective tight-binding model
for the Majorana fermions in the vortex chain.

First, let us consider an isolated vortex located at $\bs{x}=0$
in the vicinity of a domain wall of type (ii), i.e.,
at $|x_1|\ll l$, in the strongly type-II limit, 
by assuming $|\mu|\ll\Delta_0$ and $|B|\ll\Delta_0$.
We also assume that the vortex core radius,
which is given by the coherence length $\xi=v/\Delta_0$
from the analogy to conventional type-II superconductors,
is much shorter than the domain-wall thickness $l$; $\xi\ll l$.
Then, the problem is reduced to a 
Bogoliubov-de Gennes Hamiltonian
$H_{\Delta}\!\!:=\!\!
\int d^2\bs{x}\,
\Psi^\dagger(\bs{x})
\mathcal{H}_{\Delta}
\Psi(\bs{x})$,
where
\begin{eqnarray}
\mathcal{H}_{\Delta}
\!\!&:=&\!\!
\begin{pmatrix}
-\mu&vk&\Delta&0\\
vk^\dagger&-\mu&0&\Delta\\
\Delta^*&0&\mu&-vk\\
0&\Delta^*&-vk^\dagger&\mu
\end{pmatrix}.
\label{eq:BdGHamiltonian}
\end{eqnarray}
We used the bispinors 
$\Psi^\dagger(\bs{x})=
\biglb(\psi^\dagger(\bs{x}),- i\psi^{\text{T}}(\bs{x})\sigma_2\bigrb)$
and the
complex momentum operators 
$k:=-\partial_1+i\partial_2$ and $k^\dagger=\partial_1+i\partial_2$.
When $B>0$, the order parameter $\Delta(\bm{x})$
has the form
$\Delta_\pm(\bm{x})=\pm \tilde{\Delta}\left(|\bs{x}|\right)z/|\bs{x}|$
for $|\bs{x}|\ll\xi$,
where $z=x_1+ix_2$ and $\tilde{\Delta}\left(x\right)$
is a real non-negative function
with $\tilde{\Delta}(0)=0$.
With the sign $\pm$ we allow for two different choices of the gauge
for $\Delta(\bs{x})$.
A zero mode $\Psi_\pm(\bs{x})$ bound to the vortex is the solution
to
$\mathcal{H}_{\Delta_\pm}\Psi_\pm(\bs{x})=0$,
which is normalizable, i.e., 
$\int d^2\bs{x}\Psi^\dagger_\pm(\bs{x})\Psi_\pm(\bs{x})<\infty$.
To linear order in $\mu$,
it is given by
\begin{equation}
\Psi_\pm(\bs{x})
=\frac{1}{\sqrt{\mathcal{N}}}\!
\begin{pmatrix}
1\\
-\frac{\mu}{2v}z\\  
\pm \frac{\mu}{2v}z^* \\ 
\pm 1
\end{pmatrix}
\exp\!\left[-\frac{1}{v}\int_{0}^{|\bs{x}|}d x\tilde{\Delta}(x)\right]
,
\label{eq:spinor}
\end{equation}
where $\mathcal{N}$ is the normalization factor~\cite{Bergmann09}.

We turn to the case where $\Delta(\bs{x})$ includes the vortex chain
instead of an isolated vortex.
The vortices are centered at
$\bm{x}=(n-\frac12)\bm{x}_q$,
($n=0,\pm1,\pm2,\ldots$).
The overlap integral $t$ of the Majorana modes bound to two neighboring
vortices can be written in terms of two solutions for the isolated vortex,
$\Psi_+$ and $\Psi_-$, as
\begin{equation}
t:=
\int d^2\bs{x}
\Psi^\dagger_+\!\left(\bs{x}+\frac{\bs{x}_q}{2}\right)
\mathcal{H}^{\ }_{\Delta}
\Psi_-\!\left(\bs{x}-\frac{\bs{x}_q}{2}\right)
.
\label{eq:hybridization}
\end{equation}
In our case of $|B|\ll\Delta_0$, which yields $\xi\ll\pi/|q|$,
we can approximate
$\Delta=i\Delta_0$ [$\tilde{\Delta}(x)=\Delta_0$]
in Eq.~\eqref{eq:hybridization}.
We obtain 
\begin{equation}
t=C\mu\left(\frac{\Delta_0}{|B|}\right)^{3/2}
\exp\!\left(-\frac{\pi\Delta_0}{2|B|}\right)
+\mathcal{O}(\mu^2/\Delta_0)
\label{eq:t}
\end{equation}
as the leading asymptotic behavior in $|\mu|/\Delta_0\ll1$,
where $C$ is a dimensionless constant of order one that depends on
the spatial structure of $\Delta(\bs{x})$.
We note that the inclusion of the in-plane component of the Zeeman field
in Eq.~(\ref{eq:BdGHamiltonian}) for the calculation of $t$ does not
change the leading asymptotic behavior.
The above example demonstrates that we can create a chain of Majorana fermions
whose transfer integral $t$ is controlled by $\mu$ as well as $|B|/\Delta_0$.

The vanishing of the hybridization and the emergence of many Majorana
zero modes at $\mu=0$ can also be explained in terms of Weinberg's index
theorem on the number of fermion zero modes~\cite{Weinberg81,indexcomment}.
Namely, there exist at least $|n|$ zero-energy states for our Hamiltonian
Eq.~\eqref{eq:BdGHamiltonian} at $\mu=0$, where
$n:=\frac{1}{2\pi}\oint_{\mathcal{C}}d\bs{x}\cdot\nabla\,\text{arg}\,\Delta$
is the number of vortices enclosed by a loop $\mathcal{C}$.
This property survives,
when the in-plane Zeeman field $\bs{B}(\bs{x})=\bs{M}(\bs{x},0)$ 
is restored
in the Hamiltonian~\eqref{eq:BdGHamiltonian}, where $\bs{M}$ is
in the form of Eq.~\eqref{eq:B} with $j=3$.
Then, $\bs{B}$ at $|x_1|\to\infty$ can be absorbed by a gauge transformation
resulting in an exponentially decaying fictitious vector potential ${\cal A}$,
and hence the condition for the theorem~\cite{Weinberg81} is satisfied.
In contrast, if there remains a non-negligible out-of-plane Zeeman field
$B_3(\bs{x})$ unlike the above case, it couples the Majorana fermions and
lifts their degeneracy.

A tight-binding chain of Majorana fermions can be formulated in general with 
the real scalar Majorana fields $\phi_j$ by the Hamiltonian,
\begin{equation}
H_{\text{eff}}=i\sum_{j}(t+\delta t_j)\phi_j\phi_{j+1}.
\label{eq:eff theory}
\end{equation}
Here, $j$ labels the vortices along the chain 
and $\delta t_j$ are random fluctuations of the hopping amplitude
caused by spatial inhomogeneity of $\mu$.
A similar model was introduced by Kitaev in Ref.~\onlinecite{Kitaev01}
and has recently been studied in Ref.~\onlinecite{Alicea10-2}.
Since the Majorana fermions are charge neutral, they do not contribute
to the electric transport but to the thermal transport and the specific heat.
The low-temperature one-dimensional thermal conductivity $\kappa$
due to the Majorana fermions is obtained
within the semiclassical transport theory as 
$
\kappa/T=\pi^2k^2_{\text{B}}|t|\tau/(3|q|)
$.
Here $k_\mathrm{B}$ is the Boltzmann constant
and $\tau$ the elastic scattering time associated with the backscattering
due to the fluctuations $\delta t_j$.
The presence of the Majorana fermions can be confirmed by the measurement of  
their $T$-linear contribution to $\kappa$, with the characteristic
V-shaped dependence of $\kappa$ on $|\mu|$ across the Dirac nodal point.

An alternative direct way of detecting the Majorana-fermion chain
would be an STM experiment on the superconducting surface.
This requires a careful fabrication of a spatial gap in the form of
a strip separating two oppositely magnetized ferromagnets,
to allow the STM tip to probe the superconducting surface.

For these measurements, the temperature should be low enough to
suppress contributions from other midgap states bound to the vortices.
The second-lowest-energy midgap states appear at 
$|E_1|=\sqrt{3}\Delta_0/2$
in the limit $\xi|q|\to 0$ and at $\mu=0$~\cite{Ser08}.
This leads to the condition $k_{\text{B}}T\ll\Delta_0$.
The suppression of phonon and magnon contributions to $\kappa$
also requires low temperature:
The phonon contribution can be separated by its cubic temperature dependence.
An insulating ferromagnetic thin film with a large magnon gap can be used
to suppress magnon contributions.

We will now discuss a possible experimental realization of our proposal.
If a 3D TI becomes superconducting upon cooling,
the surface Dirac fermions associated with the 3D TI acquire an energy gap.
Then, this may be described as an $s$-wave superconductivity of
Dirac fermions hosting a single Dirac cone in the normal state.
When an insulating ferromagnetic thin film is deposited on this surface,
it will provide an ideal laboratory for our theoretical model.
Candidate materials for the 3D TI with superconducting instability include
Cu$_x$Bi$_2$Se$_3$ which becomes a strongly type-II
superconductor by doping Cu into the parent 3D TI Bi$_2$Se$_3$~\cite{Hor09,Wray09},
and the superconductor LaPtBi which has been predicted theoretically
as a 3D TI~\cite{Chadov10}.
The chemical potential needs to be carefully controlled by applying
a gate voltage or chemical doping.
Insulating ferromagnetic thin films of spinel compounds
CdCr$_2$Se$_4$ and CdCr$_2$S$_4$ \cite{Balitzer66} could be deposited
on the surface.
In particular, slightly off-stoichiometric compounds are useful for
inducing appreciable magnetic anisotropy in the parent
compounds~\cite{Pinch68} and thus suppressing the magnon contribution
to the thermal conductance.
A magnetic structure needs to be carefully examined
with Lorentz microscope.
Typically, the domain structures are filamentary, thus creating several 
parallel chains of Majorana fermions. Then, their contributions to $\kappa$ 
may sum up to a larger signal.
With realistic model parameters $v\approx0.3\,\mathrm{eV}\,\mathrm{nm}$
for the 3D TI Bi$_2$Se$_3$~\cite{Xia09}, 
and $\Delta_0\approx0.5\,\text{meV}$, 
the spatial separation of two neighboring vortices is given by
$\pi/q=\pi v/2B\approx5\,\mu\text{m}$, provided $B\approx0.1\,\text{meV}$.
The angle-resolved photoemission spectroscopy measurement for
Cu$_x$Bi$_2$Se$_3$ at the optimal doping $x=0.12$ shows
$\mu\approx0.5\,\mathrm{eV}$, 
which is about $0.25\,\mathrm{eV}$ above the bottom
of the 3D conduction band~\cite{Wray09}
and clearly in the limit $|\mu|\gg\Delta_0$.
In this case the transfer integral $t$ is expected to be much larger
than the value estimated from Eq.~(\ref{eq:t})
in the opposite limit, 
$|t|\approx4\times10^{-3}|\mu|$.

Lastly, we propose an alternative way of creating a chain of
Majorana fermions by using a heterostructure based on
a Rashba-semiconductor, i.e., InAs, sandwiched by an $s$-wave superconductor
and an insulating ferromagnet with a domain wall
[Fig.~\ref{fig:Heterostructure}(c)].
We apply a magnetic field parallel to $\bs{e}_3$
to tilt the magnetization deep inside the two domains of
the ferromagnetic insulator, yielding the effective Zeeman field
in the semiconductor, 
$\bs{B}\to\pm B\bs{e}_1+B_3\bs{e}_3$,
for $x_1\to\pm\infty$,
where the domain wall extends along the $\bs{e}_2$-direction.
The chemical potential needs to be tuned closed to
the $\Gamma$-point energy level, so that $B_3$ gaps out the smaller of
the two spin-orbit split Fermi surfaces.
By linearizing the dispersion near the remaining Fermi surface,
the effective noninteracting Hamiltonian takes the same form as
Eq.~\eqref{eq: dispersion}.
This can therefore produce a similar Majorana-fermion chain.
In this case, however, one cannot largely control its energy bandwidth,
since the chemical potential for the original electrons is always
away from the Dirac nodal point. 

In conclusion, we have shown that a chain of Majorana fermions can be
created along the domain wall of a ferromagnetic insulator
that is coupled to a superconductor with a single Fermi surface
of helical electrons. 
In the search for devices that allow the manipulation of Majorana fermions
as building blocks for a topological quantum computer,
the proposed structure could in principle serve as a one-dimensional
circuit path. 
In particular, the possibility to adjust the coupling
between neighboring Majorana fermions by changing the chemical potential
can be beneficial in this context.

The authors thank C. Chamon, M. Sigrist, and J. Matsuno for useful discussions.
This work was partially supported by Grants-in-Aid for Scientific Research
under Grant No.\ 19052006 (S.O.) from the MEXT of Japan,
No.\ 21740275 (S.O.) and No.\ 21540332 (A.F.) from the JSPS.

\end{document}